\documentclass[onecolumn,12pt,journal,draftclsnofoot,a4paper,oneside]{IEEEtran} 

\usepackage{amsfonts}
\usepackage{mathrsfs}
\usepackage{cite}
\usepackage{graphicx}
\usepackage{amsmath}
\usepackage{amssymb}
\usepackage{url}

\ifCLASSINFOpdf
\else
\fi

\begin{document}
\title{\huge Coalitional Games with Overlapping Coalitions for Interference Management in Small Cell Networks}
\author{\IEEEauthorblockN{Zengfeng Zhang\IEEEauthorrefmark{1}, Lingyang
Song\IEEEauthorrefmark{1}, Zhu Han\IEEEauthorrefmark{2}, and Walid
Saad\IEEEauthorrefmark{3}\\} \IEEEauthorblockA{\IEEEauthorrefmark{1}
\normalsize{School of Electronics Engineering and Computer Science, Peking University, Beijing, China}\\
\IEEEauthorrefmark{2}\normalsize{Electrical and Computer Engineering
Department, University of Houston, Houston, TX, USA} \\
\IEEEauthorrefmark{3}\normalsize{Electrical and Computer Engineering
Department,~University of Miami, FL, USA} \\
\vspace{-0.4cm}}}

\maketitle
\begin{abstract}
In this paper, we study the problem of cooperative interference
management in an OFDMA two-tier small cell network. In particular,
we propose a novel approach for allowing the small cells to
cooperate, so as to optimize their sum-rate, while cooperatively
satisfying their maximum transmit power constraints. Unlike existing
work which assumes that only disjoint groups of cooperative small
cells can emerge, we formulate the small cells' cooperation problem
as a \emph{coalition formation game with overlapping coalitions}. In
this game, each small cell base station can choose to participate in
one or more cooperative groups (or coalitions) simultaneously, so as
to optimize the tradeoff between the benefits and costs associated
with cooperation. We study the properties of the proposed
overlapping coalition formation game and we show that it exhibits
negative externalities due to interference. Then, we propose a novel
decentralized algorithm that allows the small cell base stations to
interact and self-organize into a stable overlapping coalitional
structure. Simulation results show that the proposed algorithm
results in a notable performance advantage in terms of the total
system sum-rate, relative to the noncooperative case and the
classical algorithms for coalitional games with non-overlapping
coalitions.

\end{abstract}
\IEEEpeerreviewmaketitle

\section{Introduction}
Small cell networks are seen as one of the most promising solutions
for boosting the capacity and coverage of wireless networks. The
basic idea of small cell networks is to deploy small cells, that are
serviced by plug-and-play, low-cost, low-power small cell base
stations~(SBSs) able to connect to existing backhaul technologies
(e.g., digital subscription line (DSL), cable modem, or a wireless
backhaul)~\cite{QUEK2013}. Types of small cells include
operator-deployed picocells as well as femtocells that can be
installed by end-users at home or at the office. Recently, small
cell networks have received significant attention from a number of
standardization bodies including
3GPP~\cite{QUEK2013,CHANDRASEKHAR2008,ANDREWS2012,CLAUSSEN2008,BARBIERI2012,FENG2012,FENG2013}.
The deployment of SBSs is expected to deliver high capacity wireless
access and enable new services for the mobile users while reducing
the cost of deployment on the operators. Moreover, small cell
networks are seen as a key enabler for offloading data traffic from
the main, macro-cellular network~\cite{PEREZ2009}.

The successful introduction of small cell networks is contingent on
meeting several key technical challenges, particularly, in terms of
efficient interference management and distributed resource
allocation~\cite{DENG2012,PEREZ2009,NAKAMURA2013,GRESSET2012,
GUVENC2010,CALIN2010,LIANG2012,XIE2012,ZZHANG2013}. For instance,
underlying SBSs over the existing macro-cellular networks leads to
both cross-tier interference between the macrocell base stations and
the SBSs and co-tier interference between small cells. If not
properly managed, this increased interference can consequently
affect the overall capacity of the two-tier network. There are two
types of spectrum allocation for the network operator to select. The
first type is orthogonal spectrum allocation, in which the spectrum
in the network is shared in an orthogonal way between the macrocell
and the small cell tiers~\cite{GUVENC2010}. Although cross-tier
interference can be totally eliminated using orthogonal spectrum
allocation, the associated spectrum utilization is often
inefficient~\cite{PEREZ2009}. The second type is co-channel
assignment, in which both the macrocell and the small cell tiers
share the same spectrum~\cite{CALIN2010}. As the spectrum in the
network is reused through co-channel assignment, the spectrum
efficiency can be improved compared to the case of orthogonal
spectrum allocation. However, both cross-tier interference and
co-tier interference should be considered in this case. A lot of
recent work has studied the problem of distributed resource
allocation and interference management in small
cells. 
These existing approaches include power
control~\cite{CHANDRASEKHAR2009,JO2009,LI2011,CHU2011}, fractional
frequency reuse~\cite{LEE2010}\cite{LEE2011}, interference
alignment~\cite{LERTWIRAM2012}, interference
coordination~\cite{RANGAN2012}, the use of cognitive
base-stations~\cite{ATAAR2011}\cite{HUANG2011}, and interference
cancelation~\cite{THAI2010,WILDEMEERSCH2013,ZHANGISIT2013,ZHANGASILO2013}.
In~\cite{ZHANGASILO2013}, the authors use interference cancelation
in poisson networks with arbitrary fading distribution to decode the
k-th strongest user. The authors in~\cite{WILDEMEERSCH2013}
investigated the performance of successive interference cancelation
for uplink cellular communications. The successive interference
cancelation method is also adopted in~\cite{ZHANGISIT2013} to
calculate the aggregate throughput in random wireless networks. A
distributed algorithm is proposed for minimizing the overall
transmit power in two-tier networks in~\cite{PEREZ2011}. The problem
of joint throughput optimization, spectrum allocation and access
control in two-tier femtocell networks is investigated
in~\cite{CHEUNG2012}. In~\cite{SUN2012}, the authors proposed an
auction algorithm for solving the problem of subcarrier allocation
between macrocell users and femtocell users. In~\cite{KANG2012}
and~\cite{GURUACHARYA2010}, the authors develop two Stackelberg
game-based formulations  for studying the problem of optimizing the
performance of both macrocells and femtocells while maintaining a
maximum tolerable interference constraint at the macrocell tier. The
potential of spatial multiplexing in noncooperative two-tier
networks is studied in~\cite{ADHIKARY2012}.

Most existing works have focused on distributed interference
management schemes in which the SBSs act noncooperatively. In such a
noncooperative case, each SBS accounts only for its own quality of
service while ignoring the co-tier interference it generates at
other SBSs. Here, the co-tier interference between small cells
becomes a serious problem that can significantly reduce the system
throughput, particularly in outdoor picocell deployments. To
overcome this issue, we propose to enable cooperation between SBSs
so as to perform cooperative interference management. The idea of
cooperation in small cell networks has only been studied in a
limited number of existing
work~\cite{LEEJO2011,URGAONKAR2012,GHAREHSHIRAN2013,SORET2013,PANTISANO2011}.
In~\cite{LEEJO2011}, the authors propose a cooperative resource
allocation algorithm on intercell fairness in OFDMA femtocell
networks. In~\cite{URGAONKAR2012}, an opportunistic cooperation
approach that allows femtocell users and macrocell users to
cooperate is investigated. In~\cite{GHAREHSHIRAN2013}, the authors
introduce a game-theoretic approach to deal with the resource
allocation problem of the femtocell users. In~\cite{SORET2013}, a
collaborative inter-site carrier aggregation mechanism is proposed
to improve spectrum efficiency in a LTE-Advanced heterogeneous
network with orthogonal spectrum allocation between the macrocell
and the small cell tiers. The work in~\cite{PANTISANO2011} propose a
cooperative model for femtocell spectrum sharing using a cooperative
game with transferable utility in partition form~\cite{SAAD2009}.
However, the authors assume that the formed coalitions are disjoint
and not allowed to overlap, which implies that each SBS can only
join one coalition at most. This restriction on the cooperative
abilities of the SBSs limits the rate gains from cooperation that
can be achieved by the SBSs. Moreover, the authors in
\cite{PANTISANO2011} adopt the approach of orthogonal spectrum
allocation that is inefficient on spectrum occupation for the
two-tier small cell networks.

The main contribution of this paper is to develop a novel
cooperative interference management model for small cell networks in
which the SBSs are able to participate and cooperate with multiple
coalitions depending on the associated benefit-cost tradeoff. We
adopt the approach of co-channel assignment that improves the
spectrum efficiency compared to the approach of orthogonal spectrum
allocation used in~\cite{PANTISANO2011}. We formulate the SBSs'
cooperation problem as an overlapping coalitional game and we
propose a distributed, self-organizing algorithm for performing
overlapping coalition formation. Using the proposed algorithm, the
SBSs can interact and individually decide on which coalitions to
participate in and on how much resources to use for cooperation. We
show that, as opposed to existing coalitional game models that
assume disjoint coalitions, our proposed approach enables a higher
flexibility in cooperation. We study the properties of the proposed
algorithm, and we show that it enables the SBSs to cooperate and
self-organizing into the most beneficial and stable coalitional
structure with overlapping coalitions. To our best knowledge, this
is the first work that studies overlapping coalition formation for
small cell networks. Simulation results show that the proposed
approach yields performance gains relative to both the
noncooperative case and the classical case of coalitional games with
non-overlapping coalitions such as in \cite{PANTISANO2011}.

The rest of the paper is organized as follows: In Section~II, we
present and motivate the proposed system model. In Section~III, the
SBSs' cooperation problem is formulated as an overlapping coalition
formation game and a distributed algorithm for overlapping coalition
formation is introduced. Simulation results are presented and
analyzed in Section~IV. Consequently, conclusions are drawn in
Section~V.

\section{System Model and Problem Formulation}

\begin{figure}[t!]
\centering
\includegraphics[width=4.8in]{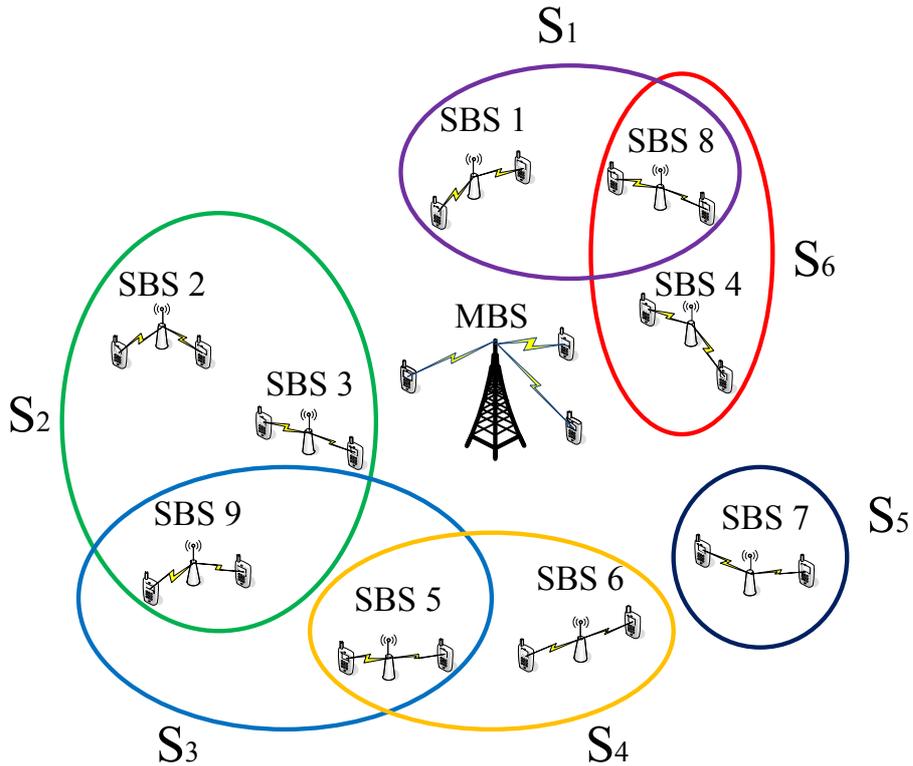}\vspace{-0.4cm}
\caption{An illustrative example of the proposed cooperative model
in small cell networks.} \label{fig:1}\vspace{-0.4cm}
\end{figure}

Consider the downlink transmission of an Orthogonal Frequency
Division Multiple Access (OFDMA) small cell network composed of $N$
SBSs and a macro-cellular network having a single macro base station
(MBS). The access method of all small cells and the macrocell is
closed access. Let $\mathcal {N} = \{1, ..., N\}$ denote the set of
all SBSs in the network. The MBS serves $W$ macrocell user
equipments (MUEs), and each SBS $i \in \mathcal {N}$ serves $L_i$
small cell user equipments (SUEs). Let $\mathcal
{L}_i=\{1,...,L_i\}$ denote the set of SUEs served by an SBS $i \in
\mathcal {N}$. Here, SBSs are connected with each other via a
wireless backhaul. Each SBS $i \in \mathcal{N}$ chooses a subchannel
set $\mathcal{T}_i$ containing $\left| {\mathcal{T}_i} \right|=M$
orthogonal frequency subchannels from a total set of subchannels
$\mathcal{T}$ in a frequency division duplexing (FDD) access mode.
The subchannel set $\mathcal{T}_i$ serves as the initial frequency
resource of SBS $i \in \mathcal {N}$. The MBS also transmits its
signal on the subchannel set $\mathcal{T}$, thus causing cross-tier
interference from MBS to the SUEs served by the SBSs. Moreover, the
SBSs are deployed in hot spot indoor large areas such as enterprises
where there are no walls not only between each SBS and its
associated SUEs, but also between all the SBSs. Meanwhile, the MBS
is located outdoor, so there exist walls between the MBS and the
SBSs.

In the traditional noncooperative scenario, each SBS $i \in \mathcal
{N}$ transmits on its own subchannels. The set of the subchannels
that SBS $i$ owns is denoted as $\mathcal{T}_i$, where
$\mathcal{T}_i \subseteq \mathcal{T}$. SBS $i$ occupies the whole
time duration of any subchannel $k \in \mathcal{T}_i$. Meanwhile,
the MBS transmits its signal to the MUEs on several subchannels from
$\mathcal{T}$, with each MUE occupying one subchannel at each time
slot. When the SBSs act noncooperatively, each SBS uses all the
subchannels from $\mathcal{T}_i$ to serve its SUEs $\mathcal{L}_i$.
For each subchannel $k \in \mathcal{T}_i$, only one SUE $u \in
\mathcal{L}_i$ is served on subchannel $k$. SUE $u$ has access to
the full time duration of subchannel $k$. We denote the channel gain
between transmitter $j$ and the receiver $u$ that owns subchannel
$k$ in SBS $i$ by $g_{j,i_u}^k$ and the downlink transmit power from
transmitter $j$ and the receiver $u$ that occupies subchannel $k$ in
SBS $i$ by $P_{j,i_u}^k$. The rate of SBS $i \in \mathcal {N}$ in
the noncooperative case is thus given by

\begin{equation}\label{eq:1}
{\Upsilon_i} = \sum\limits_{k \in {\mathcal{T}_i}} {\sum\limits_{u
\in {\mathcal{L}_i}}{{{\log }_2}\left( {1 +
\frac{{P_{i,i_u}^kg_{i,i_u}^k}}{{{\sigma ^2} + I_{MN} + I_{SN}}}}
\right)}},
\end{equation}
where ${{\sigma ^2}}$ represents the variance of the Gaussian noise,
$I_{MN} = P_{w,i_u}^kg_{w,i_u}^k$ is the cross-tier interference
from the MBS $w$ to a SUE served by SBS $i$ on subchannel $k$, and
$I_{SN}$ denotes the overall co-tier interference suffered by SUE
$u$ that is served by SBS $i$ on subchannel $k$, as
\begin{equation}\label{eq:2}
I_{SN} = \sum\limits_{j \in \mathcal {N},j \ne i}
{P_{j,i_u}^kg_{j,i_u}^k}.
\end{equation}

We note that, in dense small cell deployments, the co-tier
interference between small cells can be extremely severe which can
significantly reduce the rates achieved by the SBSs~\cite{QUEK2013}.
Nevertheless, due to the wall loss and the long distance between MBS
and SUEs, the downlink cross-tier interference is rather weak
compared to the co-tier interference between small cells. Thus, in
this work, we mainly deal with the downlink co-tier interference
suffered by the SUEs from the neighboring SBSs. In order to deal
with this interference problem, we propose a novel model in which
the SBSs are allowed to cooperate with one another as illustrated in
Fig.~\ref{fig:1}. In such a cooperative network, the SBSs can
cooperate to improve their performance and reduce co-tier
interference.

In particular, depending on the signal to noise and interference
ratio (SINR) feedbacks from their SUEs, the SBSs can decide to form
cooperative groups called \emph{coalitions} so as to mitigate the
co-tier interference between neighboring SBSs within a coalition.
The SBSs can be modeled as the players in a coalitional game. Due to
the possibility of having an SBS participating in multiple
coalitions simultaneously as shown in Fig.~\ref{fig:1}, we state the
following definition for a coalition~\cite{ZICK2012}:

\textbf{Definition 1.} A \emph{partial coalition} is given by a
vector $\mathcal{R} = \left( {{\textbf{R}_1},...,{\textbf{R}_N}}
\right)$, where $\textbf{R}_i$ is the subset of player $i$'s
resource set distributed to this coalition. The \emph{support} of a
partial coalition $\mathcal{R}$ is defined as
$\textrm{Supp}(\mathcal{R}) = \left\{ {i \in \mathcal{N}\left|
{{\textbf{R}_i} \ne \emptyset} \right.} \right\}$.

In what follows, we will omit the word ``partial'', and refer to
\emph{partial coalitions} as \emph{coalitions}.

The SBSs in the network act as players. After joining a coalition
$\mathcal{R}$, SBS $i \in \textrm{Supp}(\mathcal{R})$ allocates part
of its frequency resource into this coalition $\mathcal{R}$. Within
each coalition $\mathcal{R}$, the SBSs can jointly coordinate their
transmission so as to avoid the collisions. The resource pool of
coalition $\mathcal{R}$ can be defined as follows:
\begin{equation}\label{eq:3}
{\mathcal{T}_\mathcal{R}} = \mathop  \bigcup \limits_{{\emph{i}} \in
{\textrm{Supp}(\mathcal{R})}} {\textbf{R}_i},
\end{equation}
where $\textbf{R}_i$ denotes the subset of the frequency resources
in terms of orthogonal frequency subchannels that SBS $i \in
\textrm{Supp}(\mathcal{R})$ dedicates to the resource pool of
coalition $\mathcal{R}$ and satisfies that $\textbf{R}_i \subseteq
\mathcal{T}_i$. Here, we assume that each SBS $i$ will devote all
its frequency resources to different coalitions that it decides to
join in the network, i.e.,
\begin{equation}\label{eq:4}
\mathop  \bigcup \limits_{\{ \mathcal{R}|\emph{i} \in
{\textrm{Supp}(\mathcal{R})}\}} {\textbf{R}_i} = {\mathcal{T}_i}.
\end{equation}
Note that given SBS $i$, for any two coalitions $\mathcal{R}^t$ and
$\mathcal{R}^k$ in the network that satisfy $i \in
\textrm{Supp}(\mathcal{R}^t)$ and $i \in
\textrm{Supp}(\mathcal{R}^k)$, we have $\textbf{R}_i^t \bigcap
\textbf{R}_i^k = \emptyset$.

Without loss of generality, we consider that, whenever a coalition
$\mathcal{R}$ successfully forms, the transmissions inside
$\mathcal{R}$ will be managed by a local scheduler using the time
division multiple access (TDMA) approach as in~\cite{PANTISANO2010}.
The subchannels in ${\mathcal{T}_\mathcal{R}}$ are divided into
several time-slots. Each SBS can access only a fraction of all the
time-slots when transmitting on a specific subchannel. By doing so,
the whole superframe duration of each subchannel can be shared by
more than one SBS. Hence, the downlink transmissions from each SBS
in the coalition to its SUEs are separated. Consequently, no more
than one SBS will be using the same subchannel on the same time-slot
within a coalition, thus efficiently mitigating the interference
inside the coalition $\mathcal{R}$. However, as the resource pools
of different coalitions may not be disjoint, the system can still
suffer from inter-coalition interference. Here, we note that the
proposed approach is still applicable under any other
coalition-level interference mitigation scheme.

While cooperation can lead to significant performance benefits, it
is also often accompanied by inherent coordination costs. In
particular, for the proposed SBS cooperation model, we capture the
cost of forming coalitions via the amount of transmit power needed
to exchange information. In each coalition $\mathcal{R}$, each SBS
$i \in \textrm{Supp}(\mathcal{R})$ broadcasts its data to the other
SBSs in the coalition in order to exchange information. Here, each
SBS needs to transmit the information to the farthest SBS in the
same coalition. We assume that, during information exchange, no
transmission errors occur. So the power cost incurred for forming a
coalition $\mathcal{R}$ is given by:
\begin{equation}\label{eq:5}
P_\mathcal{R} = \sum\limits_{i \in \textrm{Supp}(\mathcal{R})}
{{P_{i,{j^*}}}},
\end{equation}
where ${P_{i,{j^*}}}$ is the power spent by SBS $i$ to broadcast the
information to the farthest SBS ${j^*}$ in a coalition
$\mathcal{R}$. Meanwhile, for every coalition $\mathcal{R}$, we
define a maximum tolerable power cost  ${P_{\lim }}$.

\section{Small Cell Cooperation as an Overlapping Coalitional Game}
In this section, we will develop an overlapping coalition formation
(OCF) game model to solve the problem of co-tier interference
management in two-tier small cell networks. OCF games have been
recently introduced in the game theory
literature~\cite{CHALKIADAKIS2010,ZICK2011,ZICK2012,ZICKMAR2012}.
OCF games have been applied in cognitive radio networks and
smartphone sensing~\cite{TWANG2013}\cite{BDI2013}. The goal is to
leverage cooperation for maximizing the system performance in terms
of sum-rate while taking into account the cooperation costs. A
distributed OCF algorithm is proposed so as to solve the developed
game. First, we will introduce some basic definitions of OCF games
in order to provide the basis of our problem solving framework.

\subsection{SBS Overlapping Coalitional Game Formulation}

We can model the cooperation problem in small cell network as an OCF
game with transferable utility in which SBSs in $\mathcal {N}$ act
as \emph{players}. To model the OCF game, we assume that each SBS
treats the subchannels it possesses as the resources that it can
distribute among the coalitions it joins. Here, we make the
following definitions~\cite{ZICK2012}:

\textbf{Definition 2.} A \emph{discrete OCF game $G = (\mathcal
{N},v)$} with a \emph{transferable utility (TU)} is defined by a set
of players $\mathcal {N}$ and a value function $v$ assigning a real
value to each \emph{coalition} $\mathcal{R}$.

We also note that $v(\emptyset) = 0$. Each player $i\in \mathcal{N}$
can join multiple coalitions. In the following, we will omit the
word ``discrete'', and refer to \emph{discrete OCF games} as
\emph{OCF games}.

\textbf{Definition 3.} An \emph{overlapping coalitional structure
over $\mathcal{N}$}, denoted as $\mathcal{CS}$, is defined as a set
$\mathcal{CS} = \left( {\mathcal{R}^1,...,\mathcal{R}^l} \right)$
where $l$ is the number of coalitions and
$\textrm{Supp}(\mathcal{R}^t) \subseteq \mathcal {N}$, $t \in
\{1,...,l\}$.

In the proposed OCF game $G = (\mathcal {N},v)$, the players, i.e.,
the SBSs $\mathcal {N} = \{1, ..., N\}$ can choose to cooperate by
forming coalitions and sharing their frequency resources. Here, we
consider, without loss of generality, that the resource pool of a
coalition $\mathcal{R}$ is divided among the SBSs in $\mathcal{R}$
using a popular criterion named \emph{proportional fairness}, i.e.,
each SBS $i \in \textrm{Supp}(\mathcal{R})$ gets an share $f_i \in
[0,1]$ of the frequency resources from the coalition $\mathcal{R}$
through the TDMA scheduling process of the proposed local scheduler,
and the share satisfies that
\begin{equation}\label{eq:7_1}
\sum\limits_{i \in \textrm{Supp}(\mathcal{R})} {{f_i}}  = 1
\end{equation}
and
\begin{equation}\label{eq:7_2}
 \frac{{{f_i}}}{{{f_j}}} =
 \frac{{{\left|\textbf{R}_i\right|}}}{{{\left|\textbf{R}_j\right|}}},
\end{equation}
where $\left|\textbf{R}_i\right|$ denotes the number of subchannels
in $\textbf{R}_i$.

The proportional fairness criterion guarantees that the SBSs that
dedicate more of its own frequency resources, i.e., subchannels to
the coalition deserve more frequency resources back from the
resource pool of the coalition. Furthermore, due to the TDMA process
within each coalition, interference inside single coalition can be
neglected, and, thus, we focus on mitigating the inter-coalition
interference.

In our model, the inter-coalition interference leads to negative
externalities, implying that the performance of the players in one
coalition is affected by the other coalitions in the network.
Therefore, the utility  $U(\mathcal{R},\mathcal{CS})$ of any
coalition $\mathcal{R} \in \mathcal {CS}$, which corresponds to the
sum-rate achieved by $\mathcal{R}$, will be dependent on not only
the members of $\mathcal{R}$ but also the coalitional structure
$\mathcal{CS}$ due to inter-coalition interference as follows:
\begin{equation}\label{eq:7}
U(\mathcal{R},\mathcal{CS}) = \sum\limits_{i \in
\textrm{Supp}(\mathcal{R})} {\sum\limits_{k \in
{\mathcal{T}_\mathcal{R}}} {\sum\limits_{u \in {\mathcal{L}_i}}
{\gamma _{i,i_u}^k{{\log }_2}\left( {1 +
\frac{{P_{i,i_u}^kg_{i,i_u}^k}}{{{\sigma ^2} + I_{MO} + I_{SO}}}}
\right)} }},
\end{equation}
where $\gamma _{i,i_u}^k$ denotes the fraction of the time duration
during which SBS $i$ transmits on channel $k$ to serve SUE $u$,
$P_{i,i_u}^k$ indicates the transmit power from SBS $i$ to its own
SUE $u$ on subchannel $k$, $g_{i,i_u}^k$ is the according channel
gain, $I_{MO} = P_{w,i_u}^kg_{w,i_u}^k$ denotes the cross-tier
interference from the MBS $w$ to SUE $u$ served by SBS $i$ on
subchannel $k$ and ${{\sigma ^2}}$ represents the noise power.
Moreover, in (\ref{eq:7}), the term $I_{SO}$ denotes the overall
co-tier interference suffered by SUE $u$ that is served by SBS $i$
on subchannel $k$ and is defined as follows:
\begin{equation}\label{eq:6}
I_{SO} = \sum\limits_{\mathcal{R}' \in \mathcal {CS}\backslash
\mathcal{R}} {\sum\limits_{j \in Supp(\mathcal{R}'),j \ne i}
{P_{j,i_u}^kg_{j,i_u}^k} },
\end{equation}
where $P_{j,i_u}^k$ and $g_{j,i_u}^k$ denote, respectively, the
downlink transmit power and the channel gain from SBS $j \in
\textrm{Supp}(\mathcal{R}')$ to the considered SUE $u$ served by SBS
$i$ on subchannel $k$.

Given the power cost of any coalition $\mathcal{R} \in \mathcal{CS}$
defined in (\ref{eq:5}), the value of coalition $\mathcal{R}$ can be
defined as follows:
\begin{equation}\label{eq:8}
v(\mathcal{R},\mathcal{CS}) = \left\{ \begin{array}{l}
U(\mathcal{R},\mathcal{CS}),\quad \mbox{if}\,\,P_\mathcal{R} \le {P_{\lim }}, \\
{\kern 1pt} \quad \,0,\quad \quad \quad \mbox{otherwise}. \\
\end{array} \right.
\end{equation}
As the utility in (\ref{eq:8}) represents a sum-rate, then the
proposed OCF game has a transferable utility (TU), since the
sum-rate can be appropriately apportioned between the coalition
members (i.e., via a proper choice of a coding strategy).

Furthermore, we can define the payoff of an SBS $i \in
\textrm{Supp}(\mathcal{R})$ as follows
\begin{equation}\label{eq:10}
{x^i}(\mathcal{R},\mathcal{CS}) = f_iv(\mathcal{R},\mathcal{CS}),
\end{equation}
where $f_i$ is the fraction of the frequency resource that SBS $i
\in \textrm{Supp}(\mathcal{R})$ gets from coalition $\mathcal{R}$.
Note that, if SBS $i \notin \textrm{Supp}(\mathcal{R})$, then we
have ${x^i}(\mathcal{R},\mathcal{CS}) = 0$.

Suppose there are $l$ coalitions in the coalitional structure
$\mathcal{CS}$. Thus, we have $\mathcal{CS} = \left(
{\mathcal{R}^1,...,\mathcal{R}^l} \right)$. An imputation for
$\mathcal{CS}$ is defined as $\textbf{x} =
\left(\textbf{x}_1,...,\textbf{x}_l\right)$, where $\textbf{x}_j =
\left({x^1}(\mathcal{R}^j,\mathcal{CS}),...,{x^N}(\mathcal{R}^j,\mathcal{CS})\right)$.
Moreover, an outcome of the game is denoted as
$\left(\mathcal{CS},\textbf{x}\right)$ with $\mathcal{CS}$ as the
coalitional structure and $\textbf{x}$ as the imputation.

Thus, the total payoff ${p^i}(\mathcal{CS})$ received by SBS $i$
from the coalitional structure $\mathcal{CS}$ is calculated as the
sum of the payoffs of SBS $i$ from all the coalitions it is
currently participating in, which is given by:
\begin{equation}\label{eq:11}
{p^i}(\mathcal{CS}) = \sum_{j=1}^{l}
{{x^i}(\mathcal{R}^j,\mathcal{CS})},
\end{equation}
where $\mathcal{CS} = \left( {\mathcal{R}^1,...,\mathcal{R}^l}
\right)$.

Consequently, in the considered OCF game, the value of the coalition
structure $\mathcal{CS}  = \left( {\mathcal{R}^1,...,\mathcal{R}^l}
\right)$ can be defined as follows
\begin{equation}\label{eq:9}
v(\mathcal{CS}) = \sum_{j=1}^{l} {v(\mathcal{R}^j,\mathcal{CS})}.
\end{equation}
Note that $v(\mathcal{CS})$ is also the system payoff.

\subsection{Properties of the Proposed Small Cells' OCF Game}
\textbf{Definition 4.} An OCF game $(\mathcal {N},v)$ with a
transferable utility (TU) is said to be \emph{superadditive} if for
any two coalitions $\mathcal{R}^1, \mathcal{R}^2 \in \mathcal{CS}$,
$v\left( {{\mathcal{R}^1} \cup {\mathcal{R}^2},\mathcal{CS}'}
\right) \ge v\left( {{\mathcal{R}^1},\mathcal{CS}} \right) + v\left(
{{\mathcal{R}^2,\mathcal{CS}}} \right)$ with ${\mathcal{R}^1} \cup
{\mathcal{R}^2} \in \mathcal{CS}'$.

\textbf{Theorem 1.} The proposed OCF game $(\mathcal {N},v)$ is
non-superadditive.

\begin{proof}
Consider two coalitions $\mathcal{R}^1 \in \mathcal{CS}$ and
$\mathcal{R}^2 \in \mathcal{CS}$ in the network with the players
 of $\textrm{Supp}(\mathcal{R}^1) \cup \textrm{Supp}(\mathcal{R}^2)$
 located far enough such that ${P_{{\mathcal{R}^1} \cup {\mathcal{R}^2}}} > {P_{\lim
 }}$. We also suppose that ${\mathcal{R}^1} \cup
{\mathcal{R}^2} \in \mathcal{CS}'$. Therefore, according to
(\ref{eq:8}), $v\left({\mathcal{R}^1} \cup
{\mathcal{R}^2},\mathcal{CS}'\right) =
 0 < v\left( {{\mathcal{R}^1},\mathcal{CS}} \right) + v\left( {{\mathcal{R}^2},\mathcal{CS}}
 \right)$. Thus, the proposed OCF game is not superadditive.
\end{proof}

This result implies that the proposed game can be classified as a
coalition formation game~\cite{SAAD2009}~\cite{HAN2011}. One of the
main features of the OCF game is that it allows different coalitions
to overlap, i.e., an SBS $i$ can simultaneously join more than one
coalition. In order to capture this overlapping feature, we allow
each SBS $i$ to divide its frequency resource into several parts,
each of which is dedicated to a distinct coalition. To better
understand this model, consider an SBS $i \in \mathcal{N}$ which is
a player in the OCF game. The initial frequency resource of SBS $i$
is the subchannel set $\mathcal{T}_i$ which is measured in
orthogonal subchannels. Here, we present the following definition:

\textbf{Definition 5.} An \emph{SBS unit} $\lambda _m^i$ is defined
as the minimum indivisible resource (or \emph{part}) of SBS $i$
which has access to a single subchannel $c(\lambda _m^i)$ from
$\mathcal{T}_i$. The number of the SBS units that SBS $i$ owns is
$M$, i.e., the number of subchannels in $\mathcal{T}_i$. We also
have $ \cup _{m = 1}^{M}c(\lambda _m^i) = {\mathcal{T}_i}$.

In the studied OCF game, if an SBS $i$ is a member of a coalition
$\mathcal{R}$, i.e., $i \in \textrm{Supp}(\mathcal{R})$, then at
least one SBS unit $\lambda _m^i$ is dedicated to coalition
$\mathcal{R}$. An SBS can decide to remove one unit from the current
coalition and dedicate it to a new coalition when such a move leads
to a preferred coalitional structure (i.e., a higher utility).

Next we discuss the stability of the solutions for the proposed OCF
game. We mainly consider a stable solution concept for OCF games
whenever only one of the players consider to reallocate a resource
unit at a time. The proposed stability concept is defined as
follows:

\textbf{Definition 6.} Given an OCF game $G = (\mathcal {N},v)$ with
a transferable utility (TU) and a player $i \in \mathcal{N}$, let
$\left(\mathcal{CS},\textbf{x}\right)$ and
$\left(\mathcal{CS}',\textbf{y}\right)$ be two outcomes of $G$ such
that $\left(\mathcal{CS}',\textbf{y}\right)$ is the resulting
outcome of one reallocation of an SBS unit of $i$ from
$\left(\mathcal{CS},\textbf{x}\right)$. We say that
$\left(\mathcal{CS}',\textbf{y}\right)$ is a \emph{profitable
deviation} of $i$ from $\left(\mathcal{CS},\textbf{x}\right)$ if the
transformation from $\left(\mathcal{CS},\textbf{x}\right)$ to
$\left(\mathcal{CS}',\textbf{y}\right)$ is feasible.

Given the above definition, we can define a stable outcome as
follows:

\textbf{Definition 7.} An outcome
$\left(\mathcal{CS},\textbf{x}\right)$ is \emph{stable} if no player
$i \in \mathcal{N}$ has a profitable deviation from it.

The corresponding coalitional structure $\mathcal{CS}$ from a stable
outcome $\left(\mathcal{CS},\textbf{x}\right)$ is a stable
coalitional structure. When no SBS units can be switched from one
coalition to another or be put alone, the coalitional structure is
stable. In order to compare two coalitional structures, we introduce
the following definition:

\textbf{Definition 8.} Given two coalitional structures $
\mathcal{CS}_P = \left( {\mathcal{R}^1,...,\mathcal{R}^p} \right)$
and $\mathcal{CS}_Q = \left( {\mathcal{C}^1,...,\mathcal{C}^q}
\right)$ which are both defined on the player set of $\mathcal {N} =
\{1, ..., N\}$,  an \emph{order} $ \succeq_i $ is defined as a
complete, reflexive and transitive binary relation over the set of
all coalitional structures that can be possibly formed.
$\mathcal{CS}_P$ is preferred to $\mathcal{CS}_Q$ for any player $i$
when $\mathcal{CS}_P \succeq_i \mathcal{CS}_Q$.

Consequently, for any player $i \in \mathcal{N}$, given two
coalitional structures $\mathcal{CS}_P$ and $\mathcal{CS}_Q$,
$\mathcal{CS}_P \succeq_i \mathcal{CS}_Q$ means that player $i$
prefers to allocate its frequency resources in the way that
$\mathcal{CS}_P$ forms over the way that $\mathcal{CS}_Q$ forms, or
at least, player $i$ prefers $\mathcal{CS}_P$ and $\mathcal{CS}_Q$
indifferently. Moreover, if we use the asymmetric counterpart of
$\succeq_i$, denoted as $\succ_i$, then $\mathcal{CS}_P \succ_i
\mathcal{CS}_Q$ indicates that player $i$ \emph{strictly} prefers to
allocate its frequency resources in the way that $\mathcal{CS}_P$
forms over the way that $\mathcal{CS}_Q$ forms.

Different types of orders can be applied to compare two coalition
structures. This includes two major categories: individual payoff
orders and coalition payoff orders. For individual payoff orders,
each player's individual payoff in the game is mainly used to
compare two coalitional structures. In contrast, for a coalition
payoff order, the payoff of the coalitions in the game is mainly
used to compare two coalitional structures. In our OCF game case, we
aim at increasing the total payoff of the coalitional structure in a
distributed way. When two coalitional structures are compared, both
the individual payoff and the coalition payoff will be considered.

When an SBS $i$ decides to change the allocation of its unit
$\lambda _m^i$ from the current coalition, it may either allocate it
to another existing coalition or make it alone, i.e., allocate it to
another completely new and independent coalition consisting of only
SBS $i$. Accordingly, we propose the following two orders to compare
two coalitional structures:

\textbf{Definition 9.} Consider a coalitional structure
$\mathcal{CS}_P = \{\mathcal{R}^1,...,\mathcal{R}^l\}$ and an SBS
unit $\lambda _m^i$ satisfying $\lambda _m^i \in \textbf{R}_i^t \in
\mathcal{R}^t$, where $t \in \{1,...,l\}$. For a coalition
$\mathcal{R}^k$ with $k \in \{1,...,l\}$ and $k \neq t$, a new
coalitional structure is defined as $\mathcal{CS}_Q =
\{\mathcal{CS}_P \backslash \{\mathcal{R}^t,\mathcal{R}^k\}\} \cup
\{\mathcal{R}^t \backslash \{\{\lambda _m^i\}\},\mathcal{R}^k \cup
\{\{\lambda _m^i\}\}\}$. In order to transform $\mathcal{CS}_P$ into
$\mathcal{CS}_Q$, $\lambda _m^i$ must be switched from the current
coalition $\mathcal{R}^t$ to another coalition $\mathcal{R}^k$.
Then, the \emph{switching order} $\triangleright_S$ is defined as
\begin{equation}\label{eq:12}
\mathcal{CS}_Q \triangleright_S \mathcal{CS}_P  \Leftrightarrow
\left\{ \begin{array}{l}
 {p^i}(\mathcal{CS}_Q) > {p^i}(\mathcal{CS}_P), \\
  v(\mathcal{CS}_Q) > v(\mathcal{CS}_P), \\
 \sum\limits_{\lambda _m^j \in \textbf{R}_j^k \in \textrm{Supp}(\mathcal{R}^k)} {{p^j}(\mathcal{CS}_Q)}  \ge \sum\limits_{\lambda _m^j \in \textbf{R}_j^k \in \textrm{Supp}(\mathcal{R}^k)} {{p^j}(\mathcal{CS}_P)}. \\
 \end{array} \right.
\end{equation}

The switching order $\triangleright_S$ indicates that three
conditions are needed when an SBS switches one of its units from one
coalition to another. These conditions are: (i) The individual
payoff of SBS $i$ is increased, (ii) the total payoff of the
coalitional structure is increased, and (iii) the payoff of the
newly formed coalition $\mathcal{R}^k$ is not decreased.

\textbf{Definition 10.} Consider a coalitional structure
$\mathcal{CS}_P = \{\mathcal{R}^1,...,\mathcal{R}^l\}$ and an SBS
unit $\lambda _m^i$ satisfying $\lambda _m^i \in \textbf{R}_i^t \in
\mathcal{R}^t$, where $t \in \{1,...,l\}$. A new coalitional
structure is defined as ${\mathcal{CS}_E} = \{\mathcal{CS}_P
\backslash \{\mathcal{R}^t\}\} \cup \{\mathcal{R}^t \backslash
\{\{\lambda _m^i\}\}\} \cup \{\{\lambda _m^i\}\}$. In order to
transition from $\mathcal{CS}_P$ to $\mathcal{CS}_E$, $\lambda _m^i$
needs to be removed from the current coalition $\mathcal{R}^t$ and
put in an independent coalition $\{\{\lambda _m^i\}\}$. Then, the
\emph{independent order} $\triangleright_I$ is defined as
\begin{equation}\label{eq:13}
\mathcal{CS}_E \triangleright_I \mathcal{CS}_P \Leftrightarrow
\left\{ \begin{array}{l}
 {p^i}(\mathcal{CS}_E) > {p^i}(\mathcal{CS}_P), \\
 v(\mathcal{CS}_E) > v(\mathcal{CS}_P). \\
 \end{array} \right.
\end{equation}

\noindent The independent order $\triangleright_I$ implies that two
conditions are needed when an SBS unit is removed from its current
coalition and made independent: (i) The individual payoff of SBS $i$
is increased and (ii) the total payoff of the coalitional structure
is increased.

Moreover, we denote $h(\lambda _m^i)$ as the history set of the SBS
unit $\lambda _m^i$. $h(\lambda _m^i)$ is a set that contains all
the coalitions that $\lambda _m^i$ was allocated to in the past. The
rationale behind the history set $h(\lambda _m^i)$ lies in that an
SBS $i$ is prevented from allocating one of its SBS units $\lambda
_m^i$ to the same coalition twice.

Using the two orders above, each SBS can make a distributed decision
to change the allocation of its units, and thus, the coalitional
structure. Moreover, the individual payoff, the coalition payoff,
and the total payoff are considered when a reallocation is
performed. For every coalitional structure $\mathcal{CS}$, the
switching order and the independent order provide a mechanism by
which the players, i.e., the SBSs can either reallocate its SBS
units from one coalition to another coalition or make its SBS units
act independently. Here, no global scheduler is required for
performing the comparisons between pairs of coalitional structures.
Furthermore, when one of the two orders is satisfied, in order to
change the coalitional structure, we need to compare the new
coalition including $\lambda _m^i$ to the coalitions in the history
set $h(\lambda _m^i)$. If the new coalition including $\lambda _m^i$
is the same with one of the previous coalition members from
$h(\lambda _m^i)$, then the reallocation of $\lambda _m^i$ cannot be
done and the coalitional structure remains unchanged. Otherwise, if
the new coalition including $\lambda _m^i$ is different from any of
the previous coalition members from $h(\lambda _m^i)$, then we
update $h(\lambda _m^i)$ by adding the new coalition into it.
Finally, to solve the proposed game, we propose a distributed
algorithm that leads to a stable coalitional structure while
significantly improving the overall network performance, as
described next.

\subsection{Proposed Algorithm for SBSs' OCF}

\vspace{+0.3cm}
\begin{table}[!t]
\renewcommand{\arraystretch}{1.3}
\caption{The Overlapping Coalition Formation Algorithm}
\label{table_1} \centering
\begin{tabular}{p{120mm}}
\hline

$\ast$ \textbf{Initial State:}

\quad The network consists of noncooperative SBSs, and the initial coalitional structrure is denoted as $\mathcal{CS} = \{\{\mathcal{T}_1\},...,\{\mathcal{T}_N\}\}$.\\

$\ast$ \textbf{Each Round of the Algorithm:}

\quad $\star$ \emph{Phase 1 - Environment Sensing:}

\quad \quad For each SBS unit $\lambda _m^i$, SBS $i$ discovers the
existing coalitions in $\mathcal {N}$.

\quad $\star$ \emph{Phase 2 - Overlapping Coalition Formation:}

\quad \quad \textbf{Repeat}

\quad \quad a) For each SBS unit $\lambda _m^i$, SBS $i$ lists the
potential coalitions that $\lambda _m^i$ may join. Suppose

\quad \quad the current coalitional structure is $\mathcal{CS}_P =
\{\mathcal{R}^1,...,\mathcal{R}^l\}$. Then there exists $l$ possible

\quad \quad coalitional structures
$\mathcal{CS}_Q^1,...,\mathcal{CS}_Q^{l - 1},\mathcal{CS}_E$.

\quad \quad b) For each SBS unit $\lambda _m^i$, SBS $i$ decides
whether to let $\lambda _m^i$ switch to another coalition

\quad \quad in the current coalitional structure based on the
switching order and the history set

\quad \quad  $h(\lambda _m^i)$. The switching order stands when
$\mathcal{CS}_Q^k \triangleright_S \mathcal{CS}_P, k \in [1,l-1]$.

\quad \quad c) For each SBS unit $\lambda _m^i$, SBS $i$ decides
whether to let $\lambda _m^i$ become independent from

\quad \quad the current coalition based on the independent order and
the history set $h(\lambda _m^i)$. The

\quad \quad independent order stands when $\mathcal{CS}_E
\triangleright_I \mathcal{CS}_P$.

\quad \quad \textbf{Until} convergence to a stable coalition
structure $\mathcal{CS}^*$.

\quad $\star$ \emph{Phase 3 - Inner-coalition Cooperative
Transmission:}

\quad \quad Scheduling information is gathered by each SBS $i \in
\textrm{Supp}(\mathcal{R})$ from its coalition
members, and transmitted within the coalition $\mathcal{R}$ afterwards.\\

\hline
\end{tabular}\vspace{-0.6cm}
\end{table}

We propose a new distributed OCF algorithm based on the switching
order and the independent order as shown in Table~\ref{table_1}.
This algorithm is mainly composed of three phases: environment
sensing, overlapping coalition formation and intra-coalition
cooperative transmission. First of all, the network is partitioned
by $|\mathcal {N}|$ single coalitions, each of which contains a
noncooperative SBS with all its SBS units. Thus, the SBSs act
noncooperatively in the beginning. Then, through environment
sensing, the SBSs can generate a list of existing coalitions in the
network~\cite{ZHAO2007,AMIRIJOO2008,ZALONIS2012}. Successively, for
each of the SBS units, the corresponding SBS decides whether to
reallocate this SBS unit based on the switching order and the
independent order. The history set of the SBS unit should also be
taken into account when the reallocation process is performed by the
SBSs. Every time a reallocation is completed, the system payoff will
be improved respectively. Note that the system payoff throughout the
paper refers to $v(\mathcal{CS})$. When no reallocation is possible,
the second phase of overlapping coalition formation terminates and a
stable overlapping coalitional structure is formed. Consequently, in
the third phase of cooperative transmission within each coalition,
the scheduling information is broadcasted from each SBS to the other
SBSs within the same coalition. In summary, the proposed overlapping
coalition formation algorithm enables the SBSs in the network to
increase their own payoff as well as the system payoff without
hurting the other members of the newly formed coalition in each
iteration and self-organize into a stable overlapping coalition
structure in a distributed way.

While the OCF game is constructed with the SBSs as the players,
every change of the coalitional structure is caused by the
reallocation of the SBS units. As the SBS units are assigned in a
distributed way by the SBSs, they seek to improve their individual
payoffs without causing the payoff of the newly formed coalition to
decrease. Meanwhile, as the goal of this work is to maximize the
system payoff in terms of sum-rate, the coalition formation process
must also consider the improvement of the total payoff of the
network. Thus, the proposed algorithm will also ensure that the
system payoff will be increased every time the coalitional structure
changes due to the reallocation of an SBS unit.

\textbf{Theorem 2.} Starting from the initial network coalitional
structure, the convergence of the overlapping coalition formation
algorithm is guaranteed.

\begin{proof}
Given the number of the SBSs and the number of the subchannels that
each SBS initially possesses, the total number of possible
coalitional structures with overlapping coalitions is finite. As
each reallocation of the SBSs' units causes a new coalitional
structure with a higher system payoff than all the old ones, the
proposed algorithm prevents the SBSs from ordering its units to form
a coalitional structure that has previously appeared. Consequently,
each reallocation of the SBSs' units will lead to a new coalitional
structure, and given the finite number of these structures, our
algorithm is guaranteed to reach a final coalitional structure with
overlapping coalitions.
\end{proof}

Next, we prove that the final coalitional structure is a stable
coalitional structure.

\textbf{Proposition 1:} Given the switching order and the
independent order, the final coalitional structure $\mathcal{CS}^*$
resulting from the overlapping coalition formation algorithm is
stable.

\begin{proof}
If the final coalitional structure $\mathcal{CS}^*$ is not stable,
then there exists a reallocation of one of the resource units of $i$
that can change the current coalitional structure $\mathcal{CS}^*$
into a new coalitional structure $\mathcal{CS}'$. Hence, SBS $i$ can
perform an operation on one of its SBS units based either on the
switching order or the independent order, which contradicts with the
fact that $\mathcal{CS}^*$ is the final coalitional structure
resulted from the convergence of the proposed OCF algorithm. Thus,
the final coalitional structure is stable.
\end{proof}

\subsection{Distributed Implementation of the OCF Algorithm}

The proposed algorithm can be implemented distributedly, since, as
explained above, the reallocation process of the SBS units can be
performed by the SBSs independently of any centralized entity.
First, for neighbor discovery, each SBS can rely on information from
the control channels which provides the needed information including
location, frequency, number of users and so on for assisting the
SBSs to cooperate and form overlapping
coalitions~\cite{ZHAO2007,AMIRIJOO2008,ZALONIS2012}. After neighbor
discovery, the SBSs seek to engage in pairwise negotiations with the
neighboring members in other coalitions. In this phase, all the
players in the network investigate the possibility of performing a
reallocation of its SBS units using the switching order and the
independent order. Finally, in the last phase, the scheduling
process is executed within each formed coalition.

Next, we investigate the complexity of the overlapping coalition
formation phase. Given a present coalitional structure
$\mathcal{CS}$, for each SBS unit from an SBS, the computational
complexity of finding its next coalition, i.e., being reallocated by
its corresponding SBS, is easily computed to be ${\rm O}\left( N
\times M \right)$ in the worse case, where $N$ is the number of the
SBSs and $M$ is the number of the SBS units that each SBS possesses.
The worst case occurs when all the SBS units are allocated in a
noncooperative way. As coalitions begin to form, the complexity of
performing a reallocation of an SBS unit becomes smaller. This is
due to the fact that when an SBS attempts to move one of the SBS
units from one coalition to another, the complexity is dependent on
the number of coalitions within the coalitional structure. Thus, the
complexity is reduced when the number of coalitions is smaller than
${N \times M}$. Furthermore, finding all feasible possible
coalitions seems to be complex at first glance. But due to the cost
of the coalition formation, the SBS networks mainly deals with small
coalitions rather than large coalitions. Moreover, both the system
payoff and the individual payoff are considered when each
reallocation of an SBS unit is performed. Thus, the constraints in
Definitions 9 and 10 reduce the number of iterations needed for
finding the final stable outcome. Consequently, for each SBS that is
willing to reallocate its SBS units, the complexity of finding the
feasible coalitions to cooperate will be reasonable.

\section{Simulation Results and Analysis}

For simulations, we consider an MBS that is located at the chosen
coordinate of $(1$~km, $1$~km$)$. The radius of the coverage area of
the MBS is 0.75~km. The number of MUEs is $10$. $N$ SBSs are
deployed randomly and uniformly within a circular area around the
MBS with a radius of $0.1$~km. There is a wall loss attenuation of
20 dB between the MBS and the SUEs, and no wall loss between the
SBSs and the SUEs. Each SBS has a circular coverage area with a
radius of 20 m. Each SBS has 4 subchannels to use and serves 4 users
as is typical for small cells~\cite{QUEK2013}. The total number of
subchannels in the considered OFDMA small cell network is 20. The
bandwidth of each subchannel is 180 kHz~\cite{QUEK2013}. The total
number of time-slots in each transmission in TDMA mode is 4. The
transmit power of each SBS is set at 20 dBm, while the transmit
power of the MBS is 35 dBm. The maximum tolerable power to form a
coalition ${P_{\lim }} = $100 dBm. The noise variance is set to
$-104$ dBm.


\begin{figure}[!t]
\centering
\includegraphics[width=4.8in]{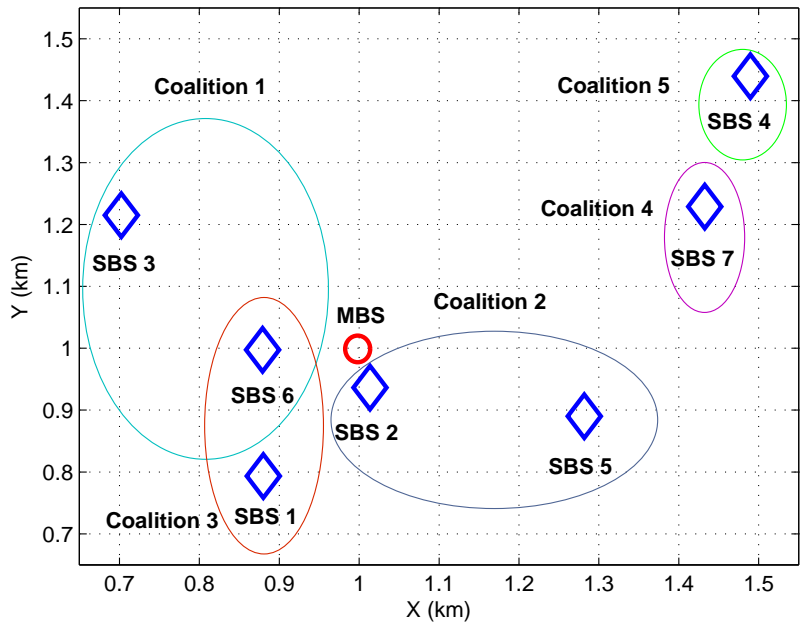}\vspace{-0.5cm}
\caption{A snapshot of an overlapping coalitional structure
resulting from the proposed approach in a small cell network.}
\label{fig:2}\vspace{-0.5cm}
\end{figure}

\begin{figure}[!t]
\centering
\includegraphics[width=4.8in]{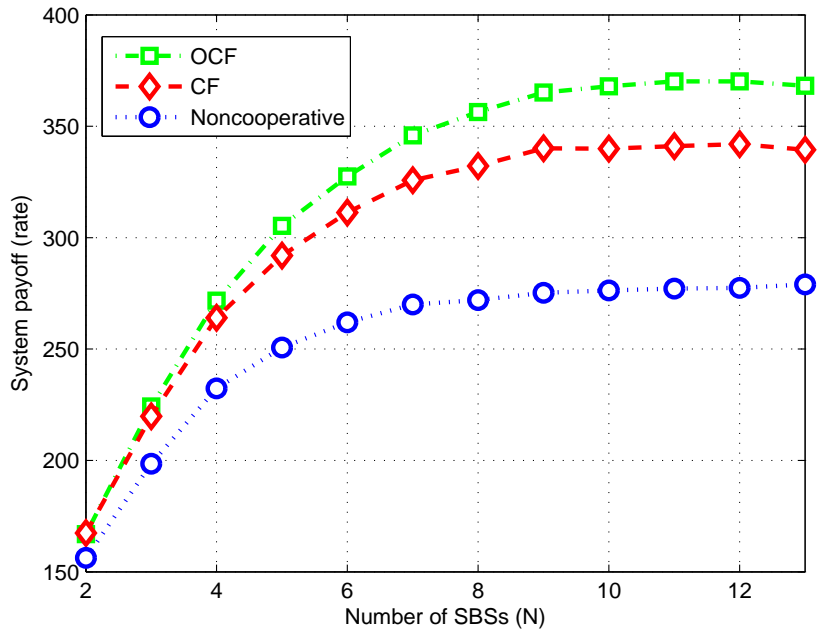}\vspace{-0.5cm}
\caption{Performance evaluation in terms of the overall system
payoff as the number of SBSs $N$ varies.}
\label{fig:3}\vspace{-0.5cm}
\end{figure}

In Fig.~\ref{fig:2}, we present a snapshot of an OFDMA small cell
network resulting from the proposed algorithm with $N=7$ SBSs. The
radius of the distribution area of SBSs is $0.7$~km. The cooperative
network shown in this figure is a stable coalitional structure
$\mathcal {CS^*}$. Initially, all the SBSs schedule their
transmissions noncooperatively. After using the proposed OCF
algorithm, they self-organize into the structure in
Fig.~\ref{fig:2}. This coalitional structure consists of 5
overlapping coalitions named Coalition 1, Coalition 2, Coalition 3,
Coalition 4, and Coalition 5. The support of Coalition 1 consists of
SBS 3 and SBS 6. The support of Coalition 2 includes SBS 2 and SBS
5. The support of Coalition 3 includes SBS 1 and SBS 6. The support
of Coalition 4 includes SBS 7. The support of Coalition 5 includes
SBS 4. SBS 4 and SBS 7 have no incentive to cooperate with other
SBSs as their spectral occupation is orthogonal to all nearby
coalitions. Meanwhile, SBS 6 is an overlapping player because its
resource units are divided into two parts assigned to different
coalitions. The interference is significantly reduced in $\mathcal
{CS^*}$ as compared to that in the noncooperative case, as the
interference between the members of the same coalition is eliminated
using proper scheduling. Clearly, Fig.~\ref{fig:2} shows that by
adopting the proposed algorithm, the SBSs can self-organize to reach
the final network structure.

Fig.~\ref{fig:3} shows the overall system utility in terms of the
total rate achieved by our proposed OCF algorithm as a function of
the number of SBSs $N$ compared with two other cases: the
non-overlapping coalition formation (CF) algorithm in
\cite{PANTISANO2011} and the noncooperative case. Fig.~\ref{fig:3}
shows that for small networks $\left(N < 4\right)$, due to the
limited choice for cooperation, the proposed OCF algorithm and the
CF algorithm have a performance that is only slightly better than
that of the noncooperative case. This indicates that the SBSs have
no incentive to cooperate in a small-sized network as the co-tier
interference remains tolerable and the cooperation possibilities are
small. As the number of SBS $N$ increases, the possibility of
cooperation for mitigating interference increases. Fig.~\ref{fig:3}
shows that, as $N$ increases, the proposed OCF algorithm exhibits
improved system performances compared to both the traditional
coalition formation game and that of the noncooperative case. The
performance advantage reaches up to $32\%$ and $9\%$ at $N=10$ SBSs
relative to the noncooperative case and the classical CF case,
respectively.

\begin{figure}[!t]
\centering
\includegraphics[width=4.8in]{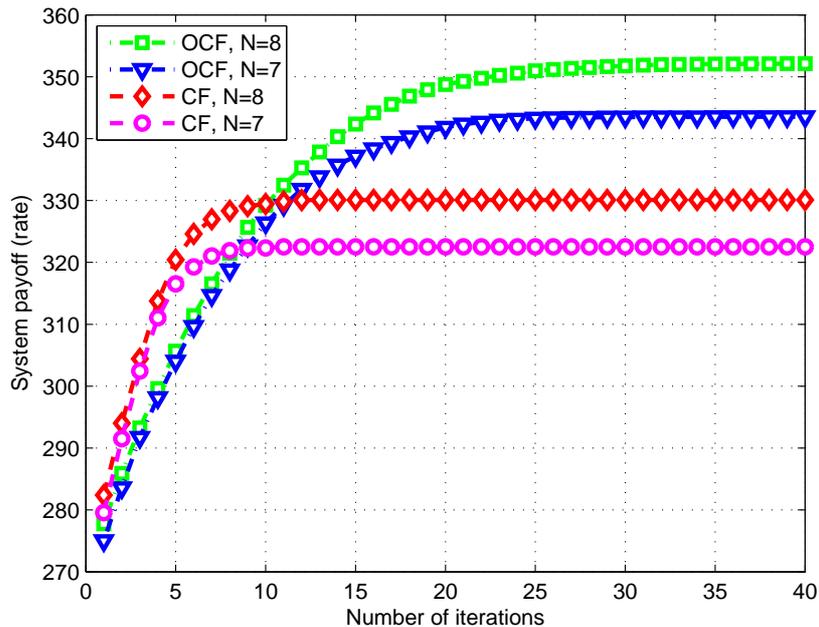}\vspace{-0.5cm}
\caption{System payoff vs. number of iterations.}
\label{fig:4}\vspace{-0.5cm}
\end{figure}

Fig.~\ref{fig:4} shows the convergence process under different
scenarios using the proposed OCF algorithm and the CF algorithm. We
observe that, although the OCF algorithm requires a few additional
iterations to reach the convergence as opposed to the CF case when
both $N=7$ and $N=8$, this number of iterations for OCF remains
reasonable. Moreover, Fig.~\ref{fig:4} shows that the OCF algorithm
clearly yields a higher system payoff than the CF case, with only
little extra overhead, in terms of the number of iterations. Hence,
the simulation results in Fig.~\ref{fig:4} clearly corroborate our
earlier analysis.

\begin{figure}[!t]
\centering
\includegraphics[width=4.8in]{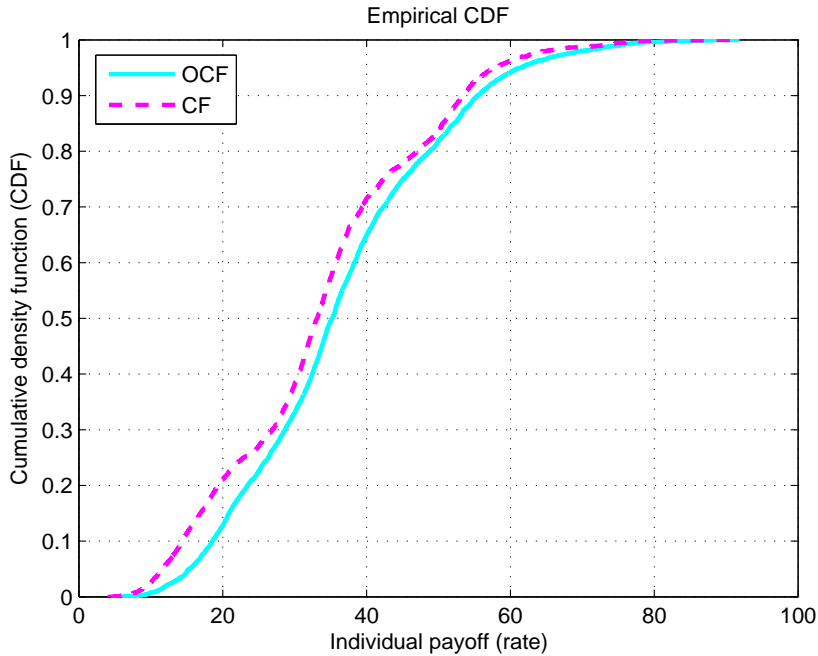}\vspace{-0.5cm}
\caption{Cumulative density function of the individual payoff for a
network with $N=10$~SBSs.} \label{fig:5}\vspace{-0.5cm}
\end{figure}

Fig.~\ref{fig:5} shows the cumulative density function (CDF) of the
individual SBS payoff resulting from the proposed OCF algorithm and
the CF algorithm when the number of SBSs is set to $N=10$. From
Fig.~\ref{fig:5}, we can clearly see that the proposed OCF algorithm
performs better than the CF algorithm in terms of the individual
payoff per SBS. For example, the expected value of the individual
payoff for a network formed from the OCF algorithm is $36$, while
for a network formed from the CF algorithm the expected value is
$33$. This is due to that our proposed algorithm allows more
flexibility for the SBSs to cooperate and form coalitions. Each SBS
is able to join multiple coalitions in a distributed way by adopting
our OCF algorithm, while it can only join one coalition at most in
the CF case. Moreover, during each reallocation, the SBSs improve
their own payoff without being detrimental to the other SBSs in the
new coalition. This also contribute to a growth of the individual
payoff of each SBS. In a nutshell, Fig.~\ref{fig:5} shows that our
proposed OCF algorithm yields an advantage on individual payoff per
SBS over the CF algorithm.

\begin{figure}[!t]
\centering
\includegraphics[width=4.8in]{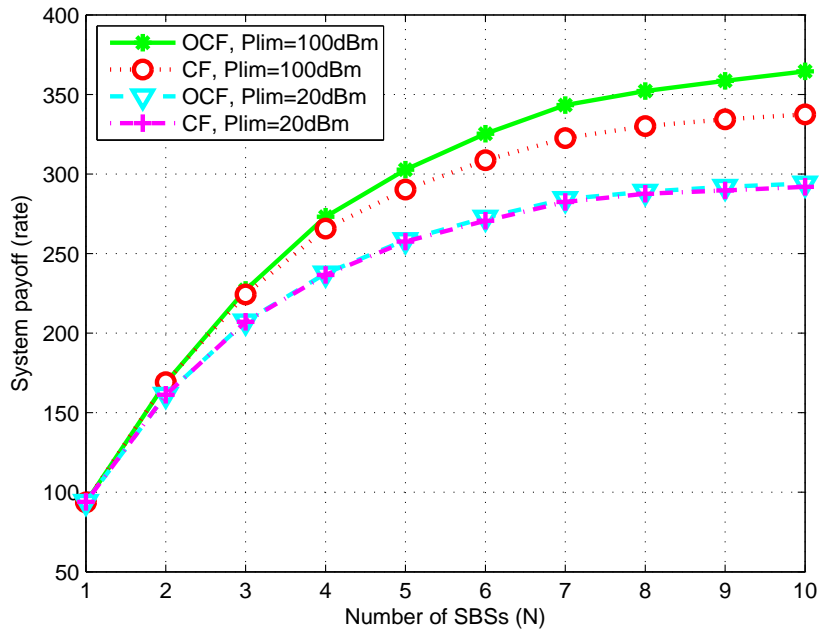}\vspace{-0.5cm}
\caption{System payoff as a function of number of SBSs $N$, for
different maximum tolerable power costs.}
\label{fig:6}\vspace{-0.5cm}
\end{figure}

Fig.~\ref{fig:6} shows the growth of the system payoff of the
network as the number of SBSs increases, under different maximum
tolerable power costs of a coalition $P_{\lim }$. Both the OCF
algorithm and the CF case are considered in Fig.~\ref{fig:6}. The
power cost incurred for forming each coalition is found from
(\ref{eq:5}). From Fig.~\ref{fig:6}, we observe that, as the number
of SBSs increases, the system payoff under two conditions both
grows. Moreover, the proposed OCF algorithm has a small advantage on
the system payoff compared to the CF case when $P_{\lim } = $20 dBm,
while the advantage of the OCF algorithm over the CF case is more
significant when $P_{\lim } = $100 dBm. This is due to the fact that
when $P_{\lim }$ is low, the SBSs can hardly cooperate with other
neighboring SBSs. Most SBSs choose to stay alone as the power cost
of possible coalitions exceeds the maximum tolerable power cost.
Thus, the system payoff of the OCF algorithm and of the CF algorithm
are close. Furthermore, when $P_{\lim }$ is high, each SBS is able
to reallocate its SBS units to join neighboring coalitions and
improve both the system payoff and its own payoff using the OCF
algorithm. Meanwhile, the cooperation possibility of the SBSs under
the CF case is also increased when $P_{\lim }$ increases.
Consequently, Fig.~\ref{fig:6} shows that the OCF algorithm incurs a
higher probability for the SBSs to cooperate than the CF case,
especially when the maximum tolerable power cost of forming a
coalition is high. Thus, our OCF algorithm achieves better system
performances in terms of sum rate than the CF algorithm.

\begin{figure}[!t]
\centering
\includegraphics[width=4.8in]{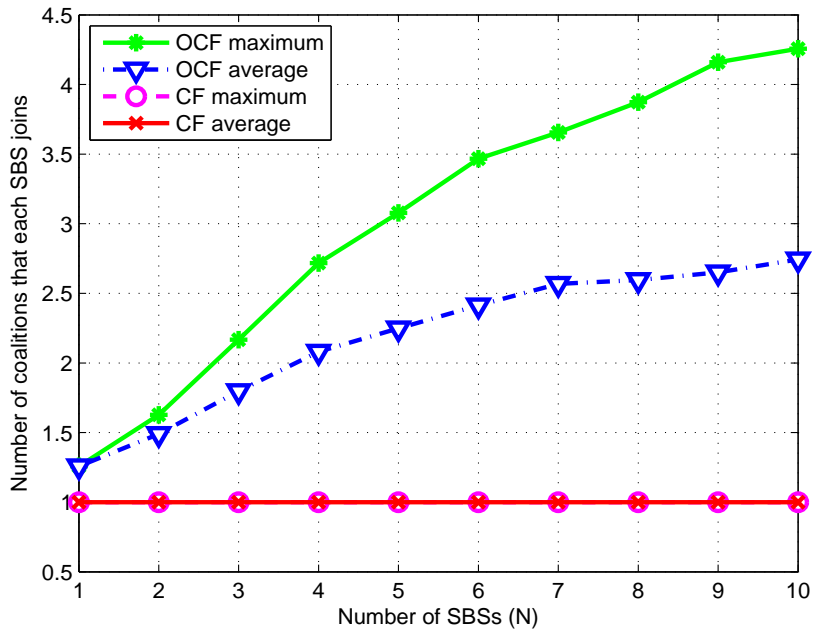}\vspace{-0.5cm}
\caption{Number of coalitions per SBS as a function of number of
SBSs $N$.} \label{fig:7}\vspace{-0.5cm}
\end{figure}

Fig.~\ref{fig:7} shows the relationship between the number of
coalitions that each SBS joins and the number of SBSs under the
proposed OCF case and the CF case. As the number of SBSs increases,
both the maximum and the average number of coalitions that each SBS
joins also grows under the OCF case. While in the CF case, each SBS
is only allowed to join one coalition at most no matter how the
number of SBSs changes, thus causing the maximum number and the
average number of coalitions that each SBS joins to remain the same
when the number of SBSs increases. Fig.~\ref{fig:7} shows that the
incentive towards cooperation for the SBSs is more significant for
the proposed OCF algorithm than for the CF case. Thus, The
cooperative gain can be achieved more efficiently by using our OCF
algorithm than the CF case when the SBSs are densely deployed in the
network. The cooperative probability of the OCF algorithm
represented by the maximum number of coalitions that each SBS joins
is $325.75\%$ larger than that of the CF case when $N=10$ SBSs are
deployed in the network.

\begin{figure}[!t]
\centering
\includegraphics[width=4.8in]{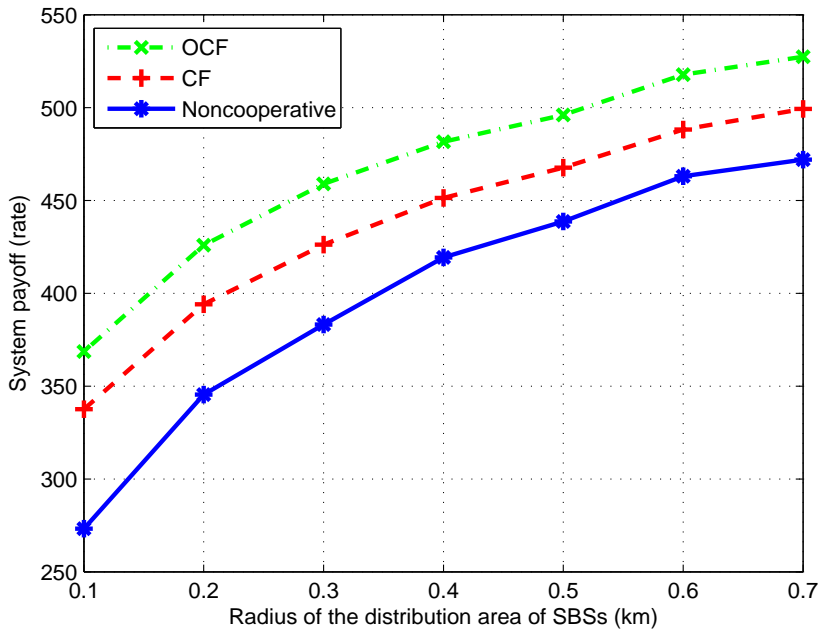}\vspace{-0.5cm}
\caption{System payoff vs. radius of the distribution area of SBSs
for a network with $N=10$~SBSs.} \label{fig:8}\vspace{-0.5cm}
\end{figure}

In Fig.~\ref{fig:8}, we show the system payoff in terms of sum-rate
as the radius of the distribution area of SBSs varies. The number of
SBSs in the network is set to $N=10$. We compare the system payoff
of the proposed OCF algorithm, CF case and noncooperative case.
Fig.~\ref{fig:8} shows that as the radius of the distribution area
of SBSs increases, the system payoff also increases. This is because
both the co-tier interference and the cross-tier interference are
mitigated when the SBSs are deployed in a larger area. Thus, the
system payoff is improved for the OCF algorithm, the CF case as well
as the noncooperative case. From Fig.~\ref{fig:8}, we can also
observe that as the radius of the distribution area of SBSs varies,
our OCF algorithm yields a higher system payoff than the CF case and
the noncooperative case.

\begin{figure}[!t]
\centering
\includegraphics[width=4.8in]{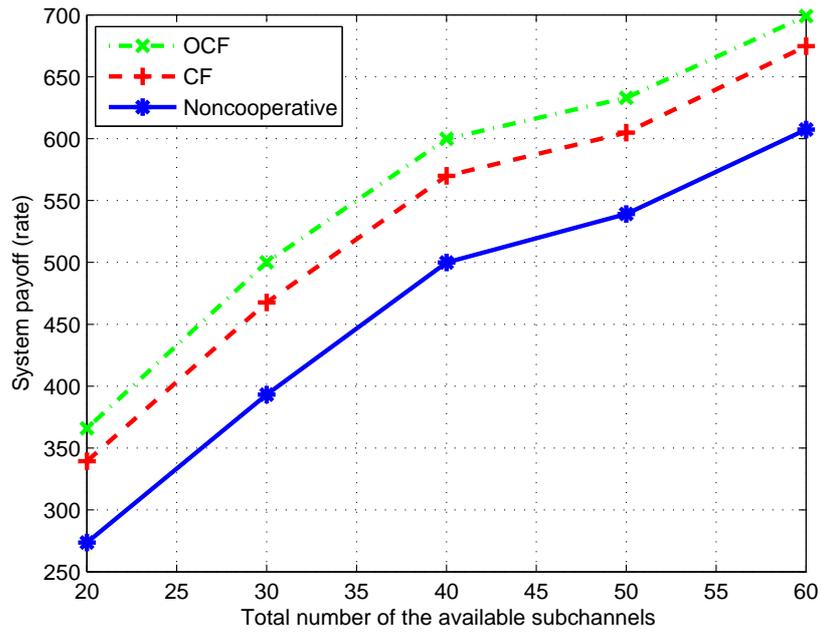}\vspace{-0.5cm}
\caption{System payoff vs. total number of subchannels for a network
with $N=10$~SBSs.} \label{fig:9}\vspace{-0.5cm}
\end{figure}

In Fig.~\ref{fig:9}, we continue to compare our OCF approach to the
CF case and the noncooperative case in terms of system payoff as the
total number of the available subchannels in the network changes.
Here, $N=10$ SBSs are deployed in the network. Note that, we adopt
the approach of co-channel assignment, i.e., the SBSs reuse the
spectrum allocated to the macrocell. Fig.~\ref{fig:9} shows that the
system payoff of the proposed OCF algorithm, the CF case, and the
noncooperative case are improved when the total number of available
subchannels increases. This is due to the fact that when the number
of available subchannels increases, the probability of conflicts on
subchannels is greatly decreased. Thus, the interference in the
two-tier small cell network is mitigated, causing the improvement of
the system payoff in terms of sum-rate. Moreover, Fig.~\ref{fig:9}
shows that our proposed OCF algorithm outperforms the CF case and
the noncooperative case in terms of system payoff when the total
number of available subchannels increases.

\begin{figure}[!t]
\centering
\includegraphics[width=4.8in]{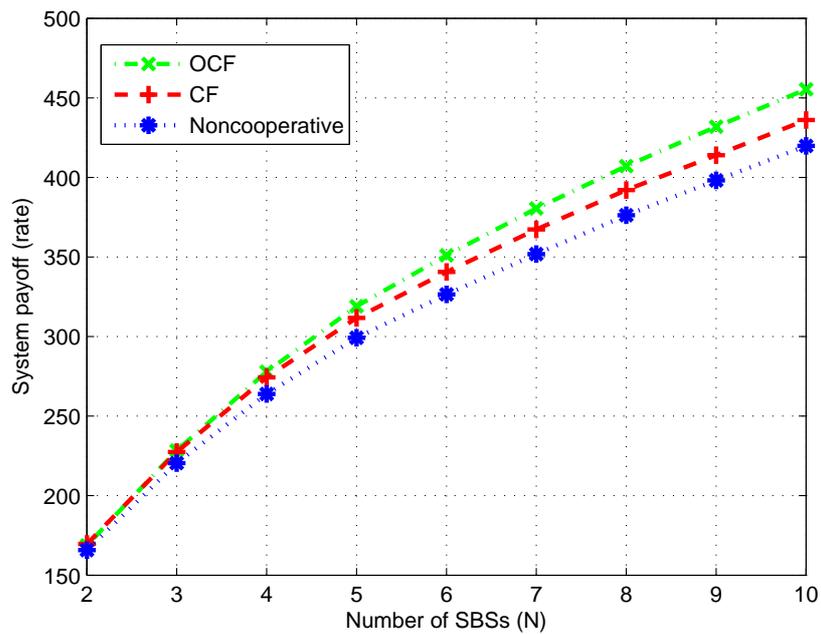}\vspace{-0.5cm}
\caption{Performance evaluation in terms of the overall system
payoff with wall loss in the small cell tier as the number of SBSs
$N$ varies.} \label{fig:10}\vspace{-0.5cm}
\end{figure}

In Fig.~\ref{fig:10}, we modify the scenario by considering the wall
loss between the MBS and the SUEs and the wall loss between the SBSs
and the SUEs, both of which are set at 20 dB. In this scenario, the
downlink cross-tier interference has a much greater impact on system
performance than in the scenario where no wall exists between the
SBSs and the SUEs such as in Fig.~\ref{fig:3}. As shown in
Fig.~\ref{fig:3} and Fig.~\ref{fig:10}, the advantage on system
payoff of our OCF algorithm over the CF algorithm and the
noncooperative case when no wall loss is considered between the SBSs
and the SUEs is more significant than that when wall loss is assumed
between the SBSs and the SUEs.

\section{Conclusions}
In this paper, we have investigated the problem of cooperative
interference management in small cell networks. We have formulated
this problem as an overlapping coalition formation game between the
small cell base stations. Then, we have shown that the proposed game
has a transferable utility and exhibits negative externalities due
to the co-tier interference between small cell base stations. To
solve this game, we have proposed a distributed overlapping
coalition formation algorithm that allows the small cell base
stations to interact and individually decide on their cooperative
decisions. By adopting the proposed algorithm, each small cell base
station can decide on the number of coalitions that it wishes to
join as well as on the resources that it allocates to each such
coalition, while optimizing the tradeoff between its overall rate
and the associated cooperative costs. We have shown that the
proposed algorithm is guaranteed to converge to a stable coalition
structure in which no small cell base station has an incentive to
reallocate its cooperative resources. Simulation results have shown
that the proposed overlapping coalitional game approach allows the
small cell base stations to self-organize into cooperative
coalitional structures while yielding notable rate gains relative to
both the noncooperative case and the classical coalition formation
algorithm with non-overlapping coalitions.


\end{document}